# 3D Atomic-Scale Metrology of Strain Relaxation and Roughness in Gate-All-Around (GAA) Transistors via Electron Ptychography

Shake Karapetyan[1], Steven E. Zeltmann[1,2], Glen Wilk[3], Ta-Kun Chen[4], Vincent D.-H. Hou[4] and David A. Muller[1,5*]

[1]School of Applied and Engineering Physics, Cornell University, Ithaca, NY, United States.

[2]Platform for the Accelerated Realization, Analysis, and Discovery of Interface Materials (PARADIM), Cornell University, Ithaca, NY, United States.

[3]Advanced Semiconductor Materials (ASM) America, Phoenix, AZ, United States.

[4]Corporate Analytical Laboratories, Taiwan Semiconductor Manufacturing Company, Hsinchu, Taiwan.

[5]Kavli Institute at Cornell for Nanoscale Science, Cornell University, Ithaca, NY, United States.

* Corresponding author: david.a.muller@cornell.edu

## Abstract

To improve transistor density and electronic performance, next-generation semiconductor devices are adopting three-dimensional architectures and feature sizes down to the few-nm regime, which require atomic-scale metrology to identify and resolve performance-limiting fabrication challenges. X-ray methods deliver three-dimensional imaging of integrated circuits but lack the spatial resolution to characterize atomic-scale features, while conventional electron microscopy offers atomic-scale imaging but limited depth information. We demonstrate how multislice electron ptychography (MEP), a computational electron microscopy technique with sub-Ångström lateral and nanometer-scale depth resolution, enables 3D imaging of buried features in devices. By performing MEP on prototype gate-all-around transistors we uncover and quantify distortions and defects at the interface of the 3D gate oxide wrapped around the channel. We find that the silicon in the 5-nm-thick channel gradually relaxes away from the interfaces, leaving only 60% of the atoms in a bulk-like structure. Quantifying the interface roughness, which was not previously possible for such small 3D structures but strongly impacts carrier mobility, we find that the top and bottom interfaces show different atomic-scale roughness profiles, reflecting their different processing conditions. By measuring 3D interface roughness simultaneously with strain relaxation and atomic-scale defects, from a single MEP dataset, we provide direct experimental values of these performance-limiting parameters needed for modeling and early fabrication optimization.

## Main Text

The increasing demands of modern computing and the drive to sustain Moore's Law have led the semiconductor industry to transition from planar to three-dimensional transistor architectures, such as Gate-All-Around (GAA) structures. This shift enables improved transistor density, performance, and power efficiency, critical for advanced integrated circuits[1]. GAA transistors, with their nanosheet-based design, provide superior electrostatic control by completely surrounding the channel with the gate electrode, unlike planar FETs or FinFETs, which lack full gate-channel coupling[1–5]. This design is ideal for ultra-scaled nodes but, as shown in Figure 1b, the intricate atomic-scale features – including interfaces, potential defects, and combinations of crystalline and amorphous materials of various composition – combined at such small length scales present significant challenges for structural and materials characterization.

Various imaging techniques have been employed to characterize semiconductor devices over the decades. Scanning Transmission electron microscopy (STEM) with electron energy loss spectroscopy, for instance, was used to establish the critical thickness of amorphous silicon oxide required for bulk-like electronic properties[6]. X-ray ptychography provides 3D imaging at deep sub-micron scales, offering insights into larger features like interconnects and the very intricate back-end elements[7]. Atom Probe Tomography (APT) enables nm-scale characterization with chemical specificity, though challenges exist due to local charging effects near insulators[8]. Atomic Electron Tomography (AET) has reported atomic-scale 3D imaging in small nanoparticles, particularly for heavier elements[9]. However, its reliance on ADF STEM limits sensitivity to light elements and introduces channeling artifacts in thicker samples, making it challenging to scale to larger, device-relevant fields of view. Conventional atomic-resolution imaging techniques, while capable of sub-Ångström lateral resolution, often fail to provide depth-resolved structural information and suffer from well-documented projection and multiple scattering artifacts[10–12].

As the critical dimensions of the GAA transistors approach the sub-10 nm regime, the limitations of existing metrologies become increasingly pronounced. Even a single defect or pinhole (a small void or discontinuity at the dielectric-semiconductor interface) can significantly impact device performance, yet such features remain difficult to detect with current techniques[1,13]. Interface roughness and strain are particularly important to quantify, as they can strongly degrade carrier mobility and shift threshold voltage[14–18]. However, there is a large variance in previously reported measurements of these values[19–21], indicating both their sensitivity to processing conditions and the limitations of conventional projective methods, which tend to underestimate the true three-dimensional nature of interface roughness. These challenges underscore the need for direct, atomic-resolution 3D measurements to more accurately characterize buried interfaces and enable predictive models of device behavior. This need has been explicitly identified in recent semiconductor roadmaps and national initiatives, including the CHIPS for America Act's Grand Challenges and review articles[1,22].

Here we show how multislice electron ptychography (MEP), a recently established four dimensional STEM (4D-STEM) technique with sub-Ångström in-plane and nanometer-scale depth resolution with sensitivity to both light and heavy atoms[23–25], bridges a critical metrology gap in device characterization highlighted above. We demonstrate that MEP enables high-fidelity three-dimensional imaging of GAA transistors, validate its accuracy on simulated structures and show experimentally that it outperforms conventional through-focal STEM in resolution and dose efficiency. Using MEP on early-stage GAA test structures from Interuniversity Microelectronics Centre (IMEC), we directly image stacking faults, pinholes, interface roughness, and strain relaxation in the crystalline-Si channel, capturing buried detrimental structural features well before electrical testing is possible. In these GAA devices, we measure the structural transition from strained interfacial silicon to bulk-like silicon to span over 2 nanometers of the 5-nanometer-wide channel (over 40%). Comparing these findings to a conformal amorphous-$SiO_2$ layer on epitaxial c-Si, we find that the strain relaxation is process-dependent and can serve as a metric for interface quality. By tracking interfacial atoms, we extract 3D maps of interface roughness amplitude and correlation length, uncovering asymmetries tied to the processing history of the interfaces. Formation of the critical GAA structure is one of the earliest steps in a CMOS process that can take 3-4 months and often exceeds 1000 steps. Early structural feedback from MEP on witness wafers can therefore accelerate development and reduce costly iterations. Moreover, since MEP captures atomic-scale interface roughness and strain, which are both key determinants of carrier mobility, it offers a direct structural metric for building predictive models of device performance and yield, enabling earlier-stage screening and optimization.

# STEM-Based 3D Imaging Techniques

Conventional STEM imaging methods can achieve three-dimensional resolution by stacking multiple two-dimensional images of the same area with shallow depth of field acquired at different defocus values, as in through-focal integrated Differential Phase Contrast (tf-iDPC), and through-focal Annular Dark Field (tf-ADF) (Figure 2c-d)[10,12,26–31]. The tf-ADF signal is generated from electrons scattered to high angles, which in very thin samples produces contrast approximately proportional to the square of the atomic number (Figure 2a blue; Figure 4c). In contrast, tf-iDPC captures the interference of low-angle scattered electrons to reconstruct the projected atomic potential, yielding in very thin samples linear or sublinear atomic number contrast and improved sensitivity to light elements (Figure 2a, purple; Figure 4b)[32].

The resolution of these methods is fundamentally constrained by the diffraction limit. The lateral Rayleigh resolution is given by $d_{xy} = 0.61\frac{\lambda}{\alpha}$, where $\lambda$ is the electron wavelength and $\alpha$ is the convergence semi-angle (Figure 2a), and depth resolution is $d_z = 2\frac{\lambda}{\alpha^2}$.[33] For a modern aberration-corrected STEM operated at typical settings of 300 kV ($\lambda$ = 1.97 pm) with $\alpha$ = 30 mrad, these correspond to a lateral resolution of 0.40 Å and a depth resolution of 44 Å. Further improvements by increasing the convergence angle are hindered by chromatic aberrations and longitudinal source coherence which degrade image quality[33,34]. Both tf-ADF and tf-iDPC suffer from systematic artifacts due to multiple scattering and channeling, especially in thicker and tilted samples, causing distortions in contrast and atomic positional accuracy[10,11,26,32]. These issues arise from complex, non-linear electron-sample interactions that are not explicitly accounted for in these imaging modes.

MEP utilizes the same setup of the microscope (Figure 2a) but overcomes these limitations by leveraging redundant information in 4D-STEM datasets to decouple the electron probe – along with its aberrations and partial coherence – from the sample's atomic potential (Figure 2b). This approach inherently accounts for multiple scattering and channeling effects by explicitly modeling the electron wave's propagation through successive slices of the sample using a multislice forward model (Figure 2b), enabling reliable 3D reconstructions of atomic potentials (Figure 2c)[24,35,36]. The development of high-speed, high-dynamic-range hybrid-pixel direct electron detectors[37,38] has enabled routine acquisition of high-fidelity 4D-STEM datasets, where a diffraction pattern is recorded at each probe position as the electron beam is raster-scanned across the sample (Figure 2a, black). The 3D information is encoded in the 4D-STEM data through parallax: the Ronchigram at each probe position is akin to a shadow image of the sample with the observer positioned at the probe crossover point. By defocusing the probe such that the crossover is around 10 nm above the sample, atoms at different depths shift across the defocused diffraction patterns at greatly different rate as the probe scans, allowing MEP to resolve their positions in depth[39].

Unlike tf-ADF and tf-iDPC (or tomography approaches), which require multiple scans, only a single 4D-STEM dataset is needed for MEP, which can reduce the total electron dose (Figure 2b-d, Extended Data Figure 3). By capturing diffraction patterns that extend well beyond the bright-field disk (Figure 2a), MEP accesses a broader range of spatial frequencies, including significant dark-field information. This enables resolution beyond the diffraction limit imposed by the probe convergence semi-angle α; under high-dose conditions, as the ones used here, the ultimate resolution instead is determined by the detector's maximum collection angle β (Figure 2a)[24,40–42].

This combination of dose efficiency, resolution, and depth sectioning has enabled 3D imaging across a wide range of materials using MEP[25,39,43–49], including our initial conference abstracts on the crystalline-Si/amorphous-SiO$_2$ interface[50,51]. The acquisition process is no more complex than that of a standard STEM image, and with recent algorithmic advances[52], initial reconstructions that once took a day can now be completed in under an hour.

## Benchmarking of the Imaging Techniques

Accurate atomic scale measurements of interface roughness and structural transitions in 3D is essential for understanding and modeling the electronic behavior of advanced semiconductor devices[14–18]. While S/TEM analysis is inherently destructive, requiring samples to be prepared from a chip portion (Figure 1a), it is standard practice in industry to dedicate test structures on the wafer for this analysis. To systematically assess imaging accuracy, we benchmarked MEP against conventional through-focal imaging methods – tf-iDPC and tf-ADF – using both simulations and experimental reconstructions.

Multislice simulations of a model rough c-Si/a-SiO$_2$/a-HfO$_2$ interface (calculated using abTEM[53]), incorporating 3D roughness and realistic experimental conditions (see Methods), show that MEP consistently resolves crystalline boundaries, interface roughness, and buried amorphous speckle contrast (Figure 3a-b and Extended Data Figure 1a-b). In contrast, tf-iDPC and tf-ADF fail to accurately recover these features (Figure 3c-d, Extended Data Figure 1c-d) due to elongation artifacts, noise, and probe-sample entanglement from channeling and multiple scattering[10,12]. MEP also has better depth resolution demonstrated by a narrower full-width at half-maximum (FWHM) in a reconstructed column (Figure 3e) and is robust to aberrations and noise. It also resolves buried features in a pMOS device model[54] – an amorphous Ta-filled pinhole at 14 nm depth – that is indistinct in tf-iDPC and only partially visible in tf-ADF, despite the latter having more than twice the dose and still achieving worse detection (Extended Data Figure 2).

We applied the same evaluation framework to a prototype IMEC GAA device (Figure 1), acquiring datasets using all three methods under identical conditions. A direct comparison of imaging capabilities is presented in Figure 4, with Extended Data Figure 3 emphasizing the improved resolution and higher signal-to-noise ratio (SNR) achieved by MEP. MEP achieves a lateral information limit of 0.49 Å, compared to 0.66 Å for tf-iDPC and 0.83 Å for tf-ADF (Extended Data Figure 3), using only half the electron dose ($0.5 \times 10^5$ e$^-$/Å$^2$ for MEP versus $1 \times 10^5$ e$^-$/Å$^2$ for tf-iDPC and tf-ADF); this enhanced resolution is limited by the small diffraction collection angle required for the ~38 nm thick section, whereas the poorer resolution of ADF and iDPC is primarily dose-limited, due to radiation damage concerns.

Figure 4a demonstrates MEP's ability to resolve depth-specific features: an intact channel at 13 nm depth (magenta) and a narrowed channel at 23 nm depth (green) from a hafnium oxide intrusion (red arrow). By contrast, tf-iDPC (Figure 4b) is affected by channeling and tilt artifacts, falsely depicting the channel as intact at both depths. Lattice details are poorly resolved, and the intrusion is hard to see, even at the higher dose. Similarly, tf-ADF images (Figure 4c) obscure silicon atoms, with poor resolution at the same electron dose as tf-iDPC. Extended Data Figure 3 further illustrates that only MEP resolves subtle structural distortions like stacking faults (blue arrows), which are not readily identifiable from the tf-ADF and tf-iDPC.

The limited depth resolution of the conventional methods also leads to ambiguity between true sample features and surface damage to the TEM lamella due to sample preparation, which can obscure structural insights and result in erroneous feature identification. For instance, a single ADF image shows an apparent hafnium oxide intrusion into the silicon channel (red arrow, Figure 1b) but cannot localize it in depth due to its projective nature, leaving uncertainty about whether this intrusion came from device processing or TEM sample preparation. MEP resolves this ambiguity:

a 3D reconstruction (Figure 1c-d) reveals the intrusion is located 10 nm beneath the lamella surface, confirming it resulted from device fabrication rather than preparation-induced surface damage. Additional experimental depth sections from a planar c-Si/a-SiO$_2$/a-HfO$_2$ stack (Extended Data Figure 4) further demonstrate MEP's ability to distinguish between intrinsic structural features and surface redeposition artifacts introduced during Focused Ion Beam (FIB) milling, reinforcing its utility for accurate 3D interface characterization.

In summary, through simulations and experiments, we demonstrate that MEP provides reliable 3D reconstructions with reduced ambiguity, outperforming tf-iDPC and tf-ADF. The latter methods suffer from contrast artifacts, poor depth accuracy, and worse dose utilization. By decoupling the electron probe shape from the atomic potential (Supplementary Figure 1), MEP enables robust imaging of structures up to 40 nm thick. Leveraging MEP's depth-resolving capabilities, we next use it to quantify atomic-scale roughness and strain in GAA structures, directly capturing buried irregularities and nanoscale deformations that can impact device performance.

## Atomic-Scale Roughness and Strain in GAAs

Building on MEP's demonstrated depth-resolving power, we now apply it to visualize and measure strain and atomic-scale roughness in Gate-All-Around (GAA) transistors – features that are critical for device performance[13–15] yet poorly captured by conventional methods[1,22]. Figure 5 shows reconstructed slices through a prototype GAA device, revealing sharp variations in channel structure at different depths. At 18 nm, the crystalline silicon channel appears largely intact, but at 8 nm and 28 nm, large sections are missing – visible as "mouse-bites" in the channel – with associated pinholes and hafnium-rich regions. The full 3D volume (Figure 5b) places these depth slices in context, while a montage of a selected *xy* section in depth (Figure 5c) shows a stacking fault defect, evidenced by a half-unit-cell lattice shift and bending of adjacent atomic columns. Additional views are shown in Extended Data Figure 5. Green arrows highlight step edges at the c-Si/a-SiO$_2$ interface, corresponding to interface roughness captured in 3D. These early-stage GAA structures were still undergoing process development, so the presence of such structural variability was expected.

To quantitatively map interface morphology and strain, we tracked the atoms in the silicon channel across depth using Atomap[55] (see Methods, Supplementary Figure 2 and Movies 1-3). By grouping atoms into bilayers from the c-Si/a-SiO$_2$ interface inwards (Figure 6a), we extracted Si–Si spacings per bilayer (Figure 6c). We find that in these GAA structures (green and gray), on average, Si–Si spacing relaxes to its bulk value over four bilayers or 11 Å from each interface. To put this into context, we compare this strain relaxation length to that of a planar and conformal a-SiO$_2$ layer on epitaxial c-Si (red), where the convergence occurs more abruptly - within just three atomic bilayers or 8 Å. Larger standard deviations in the Si-Si distances per bilayer in GAA structures (green and gray shaded regions in Figure 6c) compared to the planar interface (red) reflect a more strained and disordered interface, highlighting greater structural disorder compared to the planar interface. Given that in modern devices, including the ones studied here, the shortest direction in the Si channel is around 50 Å, with top and bottom interfaces, this means that over 40% of the silicon in the channel is under strain (Figure 6b) – an important consideration for mobility and performance.

Conventional (S)TEM approaches have long struggled to resolve buried interface roughness in 3D, leading to conflicting models: some suggesting exponential height distributions[21], while others favor Gaussian fits[19,56]. These models imply fundamentally different interface characteristics. Exponential profiles correspond to abrupt, low-correlation roughness, while Gaussian profiles indicate smoother, long-range variations, reflective of the interface's growth history. With MEP, we directly resolve buried interfaces in 3D (Figures 6d-e, Supplementary Movies 4-5). By tracking outermost atoms of the c-Si channel in 3D, we quantify roughness in these devices and the reference planar interface using standard two-parameter metrics: RMS amplitude and in-plane correlation lengths (Extended Data Figure 5). We find that smoother, intact interfaces primarily with step-edges are best described by exponentially decaying roughness (Extended Data Figure 6). In contrast, rougher interfaces, such as the one with a "mouse-bite", deviate from both exponential and Gaussian models, indicating the limits of a simple two-parameter model (Extended Data Figure 5g-h). For the top (smoother) SiGe-on-Si-derived interface, we find an RMS roughness of 2.1 Å (corresponding to two step edges) and a correlation length of 30 Å, suggesting abrupt morphology similar to that of the planar conformal interface we studied (RMS roughness of 1.7 Å and correlation length of 15 Å). The top and bottom interfaces also differ as a result of their different epitaxial histories: the SiGe-on-Si bottom interfaces tend to have "mouse-bites" and pinholes, while the top Si-on-SiGe interface is smoother, consistent with strain-driven roughening and Si interdiffusion[57–59]. "Mouse-bites" are present in some of the top interfaces too, but less frequently.

Spatially-resolved roughness analysis also complements our simultaneous strain mapping within the same volumetric dataset, enabling the decoupling of strain and morphology. The variations we found show geometry and process-dependent differences in c-Si/a-SiO$_2$ interface quality, which we can now directly measure using MEP. This combined metrology is especially crucial at extreme scaling, where roughness and strain strongly influence channel carrier mobility, threshold voltage, and device reliability. Indeed, recent studies on sub-5 nm silicon nanoribbons attribute performance degradation to precisely these effects[4], which we can now measure directly from a single dataset, opening a path towards more studies to understand these parameters better for more predictive and realistic modeling. Interface roughness, and the need to measure it directly in buried interfaces in 3D is also relevant for quantum devices, where interface disorder leads to charge noise and decoherence, as recently demonstrated for Si/SiGe spin qubits[56].

Importantly, for the information gained, MEP is relatively time- and resource-efficient: from sample preparation to a fully reconstructed 3D image takes only about two days, providing rapid atomic-scale structural feedback. MEP uses the same sample preparation and geometry as standard STEM imaging and is performed on the same instrument, with only the 4D-STEM detector as an additional hardware requirement. MEP can be done after any fabrication process, well before electrical testing, that would require months of fabrication to produce a testable device, is possible, offering insight into potential failure modes. In fact, with experience and recently published packages, like PtyRAD[52], a preliminary 3D reconstruction of a ~20×20×30 nm volume was done in under an hour. The examined GAA structures were in the early stages of development, with synthesis still non-optimized; as such, the presence of severe defects and interface roughness was expected. Fully-processed, commercially available production devices have also been examined by the Cornell group and do not exhibit these defects or this level of roughness, indicating that optimized processing can eliminate such issues and that structural defects present in prototype structures are absent in fully optimized commercial devices.

## Conclusions

This study demonstrated how MEP provides atomic-scale 3D access to device performance limiting features, namely interface roughness and strain. Through simulation comparisons and experimental validation, we showed that MEP surpasses conventional STEM imaging techniques in recovering atomic-scale features with sub-Ångström lateral and few-nanometer depth resolution. We measured depth-resolved interface roughness and imaged stacking faults in crystalline silicon channels in prototype GAA structures. We found that the smoother, top interface (formerly SiGe-on-Si) primarily has step-edge-driven roughness which follows an exponential spatial correlation, while the rougher bottom interface (formerly Si-on-SiGe) has more pinholes and exhibits irregular roughness fluctuations – likely tied to its epitaxial growth history. In these same devices, we also found that silicon reaches its bulk-like structure only after 11 Å from the c-Si/a-SiO$_2$ interface – over 40% of the channel height – compared to 8 Å in a control planar interface, quantifying process-dependent interface quality. By enabling precise, spatially resolved quantification of these parameters inside device structures, not resolvable with standard imaging, MEP provides a powerful framework for assessing interface quality across both classical and quantum devices. As advanced devices approach atom-scale dimensions, these capabilities offer a path to close metrology gaps highlighted in the CHIPS Advanced Metrology roadmap, and pave the way for more predictive design, early defect detection, and a deeper understanding of process-structure-performance relationships at the atomic scale.

## Acknowledgements

S.K., D.A.M. acknowledge funding from TSMC through a JDP. S.E.Z. acknowledges funding from the Platform for the Accelerated Realization, Analysis, and Discovery of Interface Materials (PARADIM), which is supported by the National Science Foundation under Cooperative Agreement No. DMR-2039380. This work made use of the electron microscopy facility of PARADIM and Cornell Center for Materials Research shared instrumentation facility with Helios FIB supported by NSF (DMR-1539918). The authors also thank Malcolm Thomas, Mariena Silvestry Ramos, Philip Carubia, and John Grazul for technical support and maintenance of the electron microscopy facilities. The authors gratefully acknowledge Michael Givens (ASM), Naoto Horiguchi (imec), Hans Mertens (imec), and Hiroaki Arimura (imec) for providing the Gate-All-Around (GAA) sample used in this study. We thank Eurofins Nanolab Technologies for preparing the GAA TEM lamella used in this study. We thank Frieder Baumann for the c-Si/a-SiO2 structural model, Richard Aveyard and Bernd Rieger for the pMOS structural model. S.K. gratefully acknowledges Harikrishnan K.P., Ariana Ray, and Salva Rezaie for training and tutorials on MEP, Dasol Yoon for insightful discussions on simulations, Xiyue Zheng for sharing the automated DPC acquisition code, and Yi Jiang for helpful discussions about MEP.

## Author Contributions

The research plan was formulated by S.K., T.K.C, V.D.H.H and D.A.M. G.W. sourced the GAA structures. S.K. performed STEM and MEP characterization (experimental and simulation) under the supervision of S.E.Z. and D.A.M. S.K wrote the paper with feedback from all authors. All authors discussed the results and commented on the paper.

## Ethics declarations

Cornell University (D.A.M.) has licensed the EMPAD hardware to Thermo Fisher Scientific. The other authors declare no competing interests.

## Data and code availability

All data and codes used will be made available upon publication.

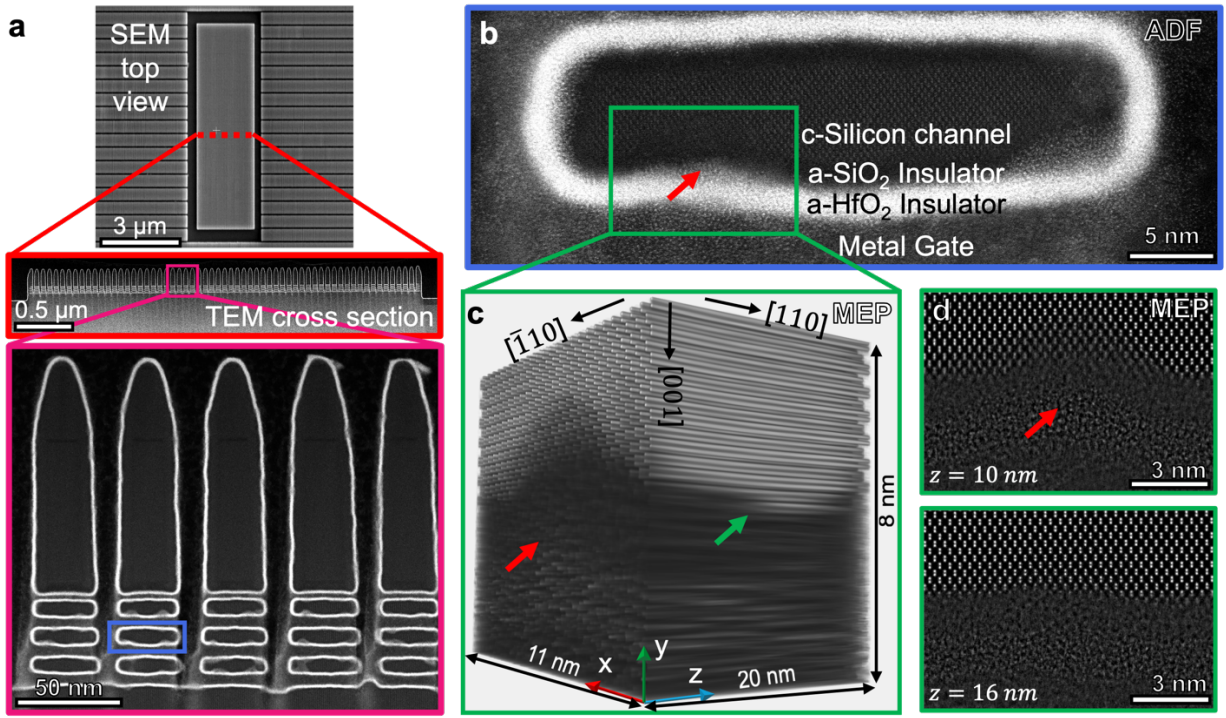

**Figure 1. Overview of the GAA structure, spanning details from microns to Ångströms, with depth-resolved imaging enabled by multislice electron ptychography (MEP). a** Top: SEM top view of the IMEC GAA node (wafer from a series of test structures for early process development, expected to have many defects) with a red dotted line indicating the region where a depth-wise electron-transparent cross-section (lamella) was prepared. Center: Wide field of view ADF image of the indicated cross-section with 63 columns of GAA structure (3 per column). Bottom: Magnified view of the pink square of the previous panel, highlighting 5 GAA columns. A single GAA structure is highlighted in blue and zoomed in on in **b**. **b** ADF image of an isolated GAA structure with labeled components; the lighter silicon and oxygen atoms are less visible due to ADF's atomic number contrast. Hafnium oxide intrusions into the silicon channel are visible, one of them highlighted with a red arrow: it is unclear if the intrusion extends throughout the entire depth of the interface and is a part of the structure or if it is fully localized at the top of the lamella and is due to sample preparation. The green-marked region was imaged by MEP, with results shown in **c-d**. **c** 3D rendering of the MEP-reconstructed region from b and d, capturing interface roughness in three dimensions with the hafnium oxide intrusion (pinhole) highlighted with a red arrow, and a step edge at the c-Si/a-SiO$_2$ interface highlighted with a green arrow. These structural details, important for device performance, while readily apparent with MEP, can be inaccessible or easily missed with conventional imaging methods. **d** MEP reconstructed images of a segment of the GAA structure highlighted in a green rectangle in **b** and shown in **c**: due to linear atomic number contrast, all sections of the structure are visible. Using MEP, we see that the hafnium oxide intrusion (top, z = 10nm, red arrow) is localized in depth and does not extend through the entire interface (bottom, z = 16nm).

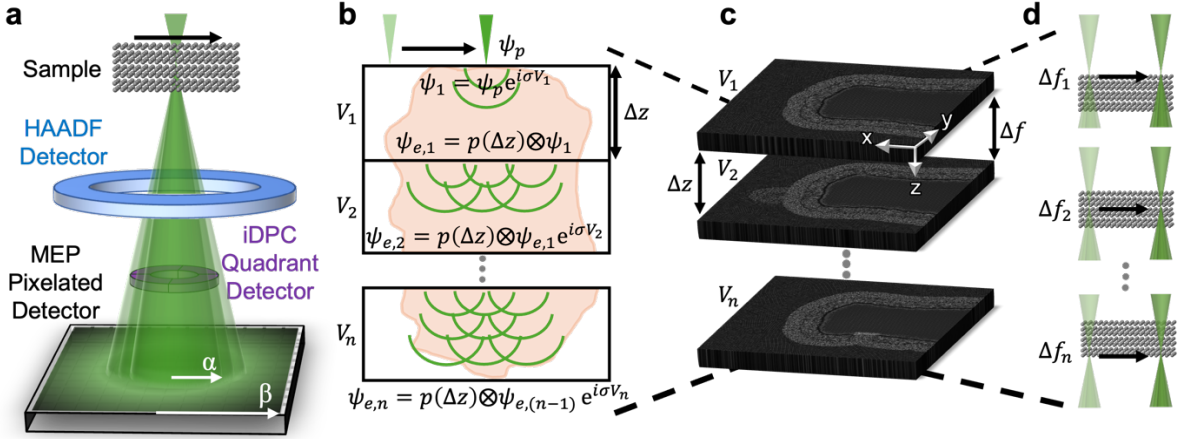

**Figure 2. Experimental setup of MEP, tf-iDPC, and tf-ADF for 3D imaging. a** Outline of the experimental setup in the STEM, where a converged probe with convergence angle α (green) is raster-scanned over an electron-transparent sample (gray). Different imaging modes are achieved by collecting transmitted electrons using various detectors ranging from a single pixeled ADF detector (blue) that collects high-angle scattered electrons, to a quadrant detector (purple, 4-8 pixels) that collects low-angle transmitted electrons for iDPC imaging and an EMPAD (black, $128^2$ pixels) that records most of the transmitted electrons, with outer collection angle β. **b** Illustration of the multislice electron ptychography (MEP) reconstruction process, where the probe is untangled from the sample's potential recovering the atomic potential V with lateral and depth resolution based on the detector collection angle β shown in **a**. For each probe position (example in green: $\psi_p$), the atomic potential of the sample is iteratively reconstructed from recorded 4D-STEM data. In the process, the sample is divided into multiple slices of thickness Δz along the propagation direction of the probe, accounting for multiple scattering of the electrons in the sample, resulting in a 3D atomic potential with depth-resolved information. **c** Visualization of depth-resolved slices obtained by different methods, where MEP slices represent reconstructed atomic potentials $V_i$ at depths *z* (z-slices) and through-focal (tf) series represent images (of probe convolved with potential) at a range of defocus values *f* (*f*-slices). **d** Schematic of tf-iDPC and tf-ADF experiments, where 2D images are acquired at different defocus values *f*. These 2D slices are aligned to construct the 3D structure of the sample, convolved with the 3D point-spread function of the probe. tf-iDPC and tf-ADF neglect changes to the probe from multiple scattering and channeling in the sample, resulting in artifacts.

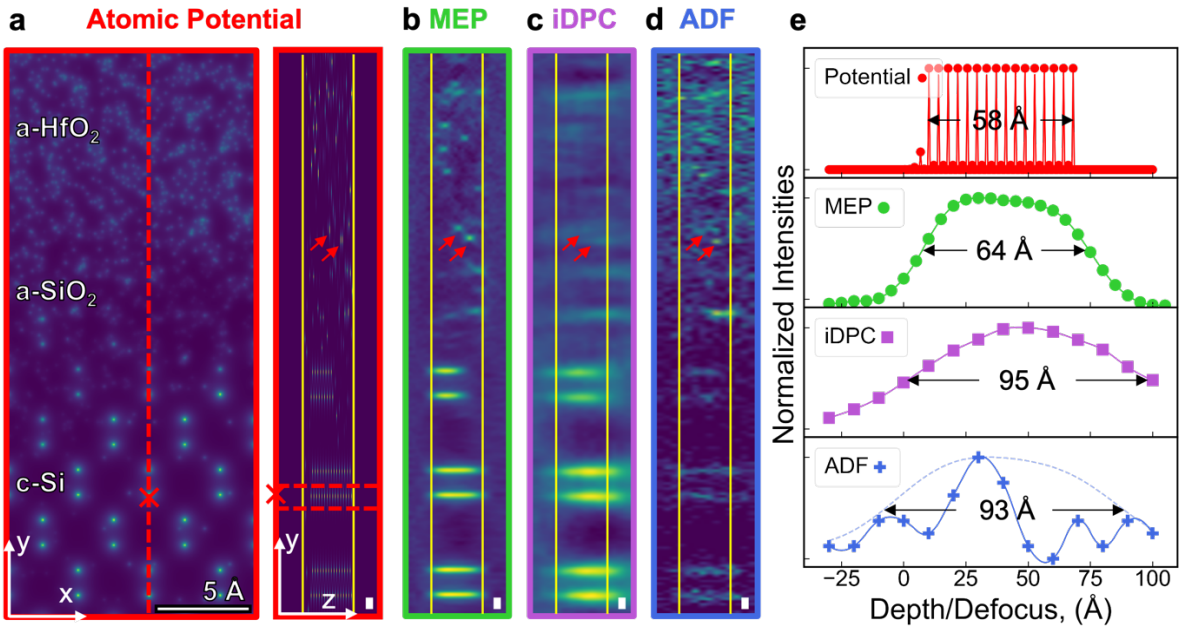

**Figure 3. Benchmarking depth-sectioning capabilities of MEP, tf-iDPC, and tf-ADF imaging techniques using a model c-Si/a-SiO$_2$/a-HfO$_2$ interface structure. a** Left: Projection of the atomic potential for a rough c-Si/a-SiO$_2$/a-HfO$_2$ interface model. Right: Depth section along the red dashed line, highlighting a Si column ("x" and red dashed-lines), and marked single (left) and double (right) a-HfO$_2$ particles with red arrows. Yellow vertical lines indicate the position and extent of the ground truth atomic model in all depth sections, showing entrance and exit interfaces. **b-d** Simulated depth reconstructions along the same line as in **a** using MEP, tf-iDPC, and tf-ADF. The simulations account for realistic spherical and chromatic aberrations, with Poisson noise added to achieve a total dose of $2.5 \times 10^5$ e$^-$/Å$^2$, emulating typical experimental conditions. Scale bars: 10 Å (unannotated). **b** MEP depth slices show accurate relative positions of all interface components and precise depth extent of the model, including clearly distinguishable entrance and exit surfaces (ground truth marked with vertical yellow lines). The single and double a-HfO$_2$ particles (red arrows) are distinguishable despite their slight elongation in depth. **c** tf-iDPC depth series shows significant elongation of all interface components in depth. The entrance and exit surfaces (ground truth marked with vertical yellow lines) are difficult to resolve and shifted, and the a-HfO$_2$ particles are smeared, rendering them indistinguishable. **d** tf-ADF depth series, where the noisy signal obscure depth features, making it challenging to differentiate interface components or resolve entrance and exit surfaces (ground truth marked with vertical yellow lines). The a-HfO$_2$ particles (red arrows) are indiscernible from noise. **e** Intensity profiles along the column marked with a red "x" in **a**. The top plot shows the actual extent of the Si column of the atomic model. The following plots show the intensity profiles of that column with different imaging modes with the annotated full-width at half-maximum (FWHM) values for each imaging mode, with MEP (green circles) demonstrating better resolution and depth accuracy compared to tf-iDPC (purple squares) and tf-ADF (blue crosses). The noise-free tf-ADF signal along this column (blue dotted line)

reveals intrinsic depth elongation of Si columns, even without noise. **Extended Data Figure 1** provides selected depth slices comparing the three imaging techniques.

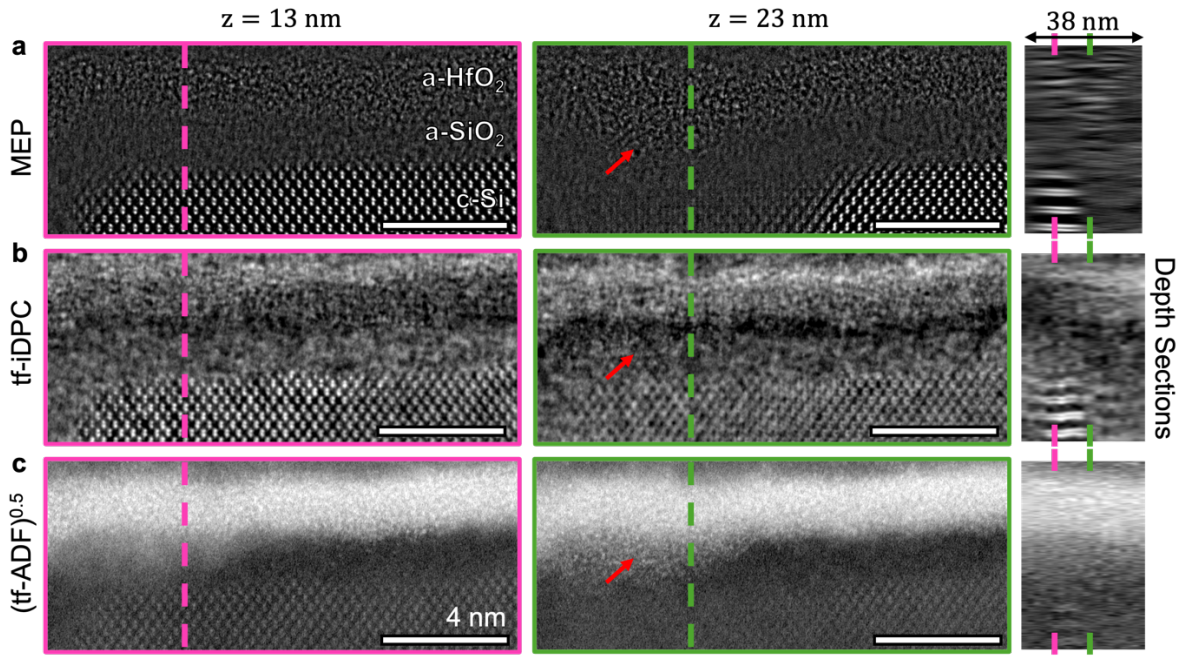

**Figure 4. Experimental comparison of depth-specific imaging capabilities of MEP, tf-iDPC, and tf-ADF for the same GAA region. a** MEP reconstructed atomic potential of the GAA region reveals an intact channel at 13 nm depth (magenta) and a partially missing channel at 23 nm depth (green), where a hafnium oxide intrusion is clearly visible (red arrow). Total dose: $0.5 \times 10^5$ e$^-$/Å$^2$. **b** tf-iDPC images of the same features at corresponding defocus values. At 13 nm, the channel also appears intact, but at 23 nm, the channel erroneously seems intact, with poorly resolved lattice details and no clear indication of the hafnium oxide intrusion (red arrow). Total dose: $1 \times 10^5$ e$^-$/Å$^2$. **c** tf-ADF images, displayed as the square root of the raw data to enhance contrast for heavy and light atoms. Silicon atoms are largely obscured due to dose-limited contrast, primarily showing the hafnium oxide and its intrusion at 23 nm defocus (red arrow). Total dose: $1 \times 10^5$ e$^-$/Å$^2$. A comparison of defects captured by these methods, along with additional results, is presented in **Extended Data Figure 3**.

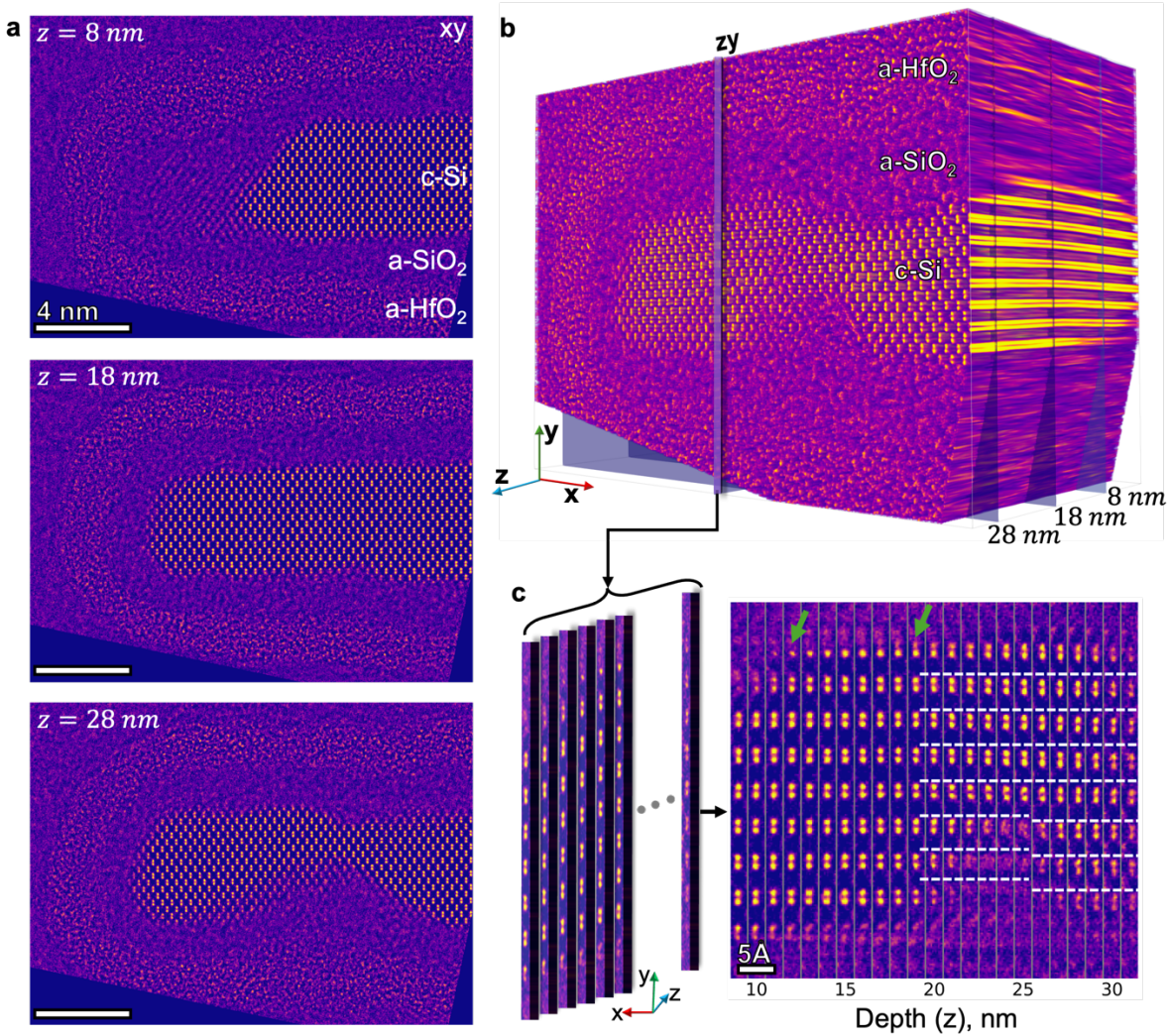

**Figure 5. Buried defects and rough interfaces in GAA structure 1 identified using MEP. a** MEP reconstruction of GAA structure 1 at different depths reveals an intact channel at z=18 nm (near the center) but irregularities at 8 nm and 28 nm, where sections of the crystalline channel are missing. **b** 3D cuboid representation of the GAA reconstruction, illustrating all sliced planes with labeled components visible in three dimensions. The cuboid dimensions are 18 nm × 13 nm × 33 nm (not to scale in depth). **c** Schematic of the depth section series along the silicon column highlighted with the zy plane in **b** (left) and a montage of those slices (right). Step edges at the c-Si/a-SiO$_2$ interface are marked with green arrows. The region with missing crystalline Si is accompanied by a stacking fault, outlined with white dotted lines that have a mismatch at the stacking fault. Additional depth slices showing more stacking faults are shown in **Extended Data Figure 4**. The Si channel shape and interface roughness of this structure are detailed in **Extended Data Figure 5 d–f**.

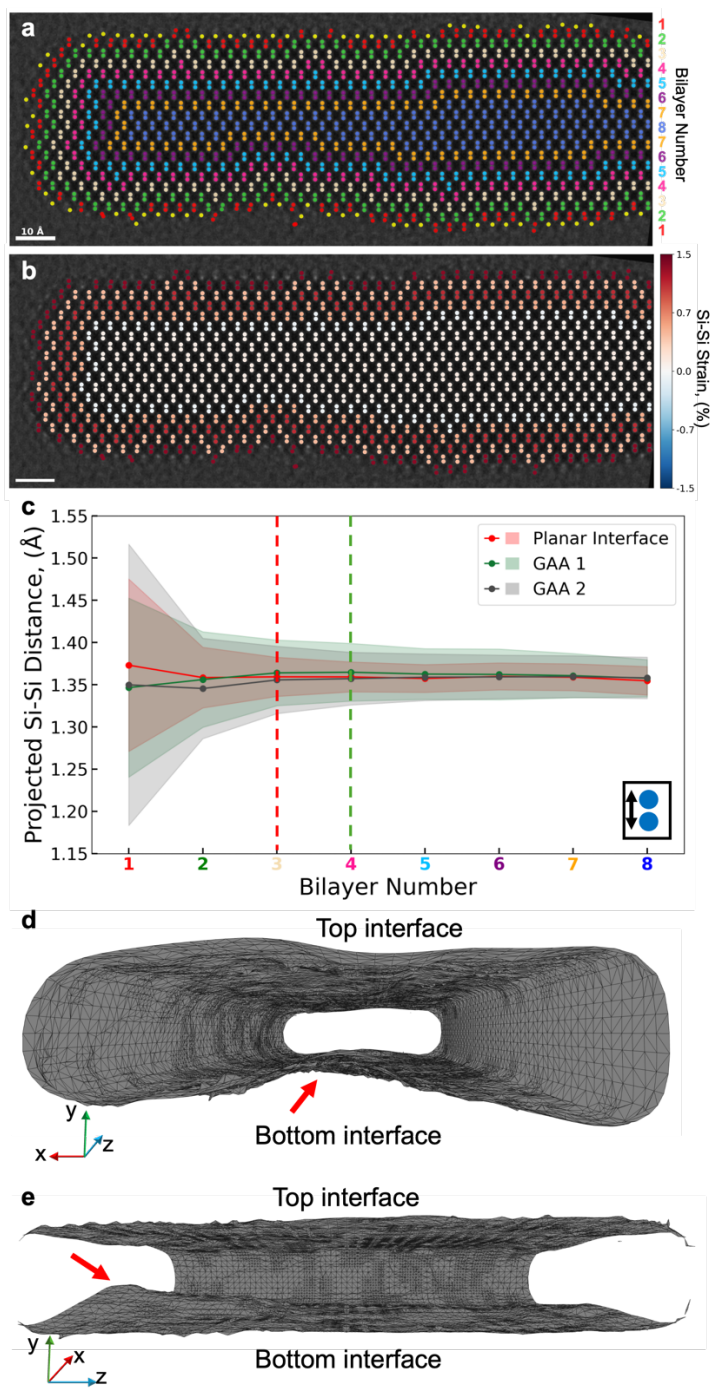

**Figure 6. Strain transition along with interface-to-bulk transition and 3D shape of a GAA structure. a** Tracked atoms overlaid on a depth slice (z=14nm), with layer-numbering convention for Si bilayers. Colors indicate each pair's relative distance from the c-Si/a-SiO$_2$ interface, where red denotes the interfacial layer. Single detected atoms are shown in yellow. **b** Recolored slice from **a**, with layers colored based on the average Si–Si distance (converted to strain) at that depth. The strained interfacial region (first eight Si atoms or four bilayers) highlights the strain transition from the interface. **c** Overlay of projected Si–Si distances as a function of layer number from a planar interface and GAA structures. The x-axis is colored per the layer scheme in **a**. The planar epitaxial c-Si/a-SiO$_2$ interface (red) converges to bulk-like Si–Si distances within three atomic pairs, while the GAA interface requires four pairs, with greater standard deviation, indicating a more strained interface in these GAA structures. **d** 3D rendering of GAA 2's surface, showing variations in interface roughness in depth and a significant intrusion at the bottom interface (red arrow). **e** Perpendicular view of **d**, providing an alternate perspective of the silicon channel surface morphology, highlighting the missing section (red arrow). Depth-resolved interface roughness is detailed in **Extended Data Figure 5**.

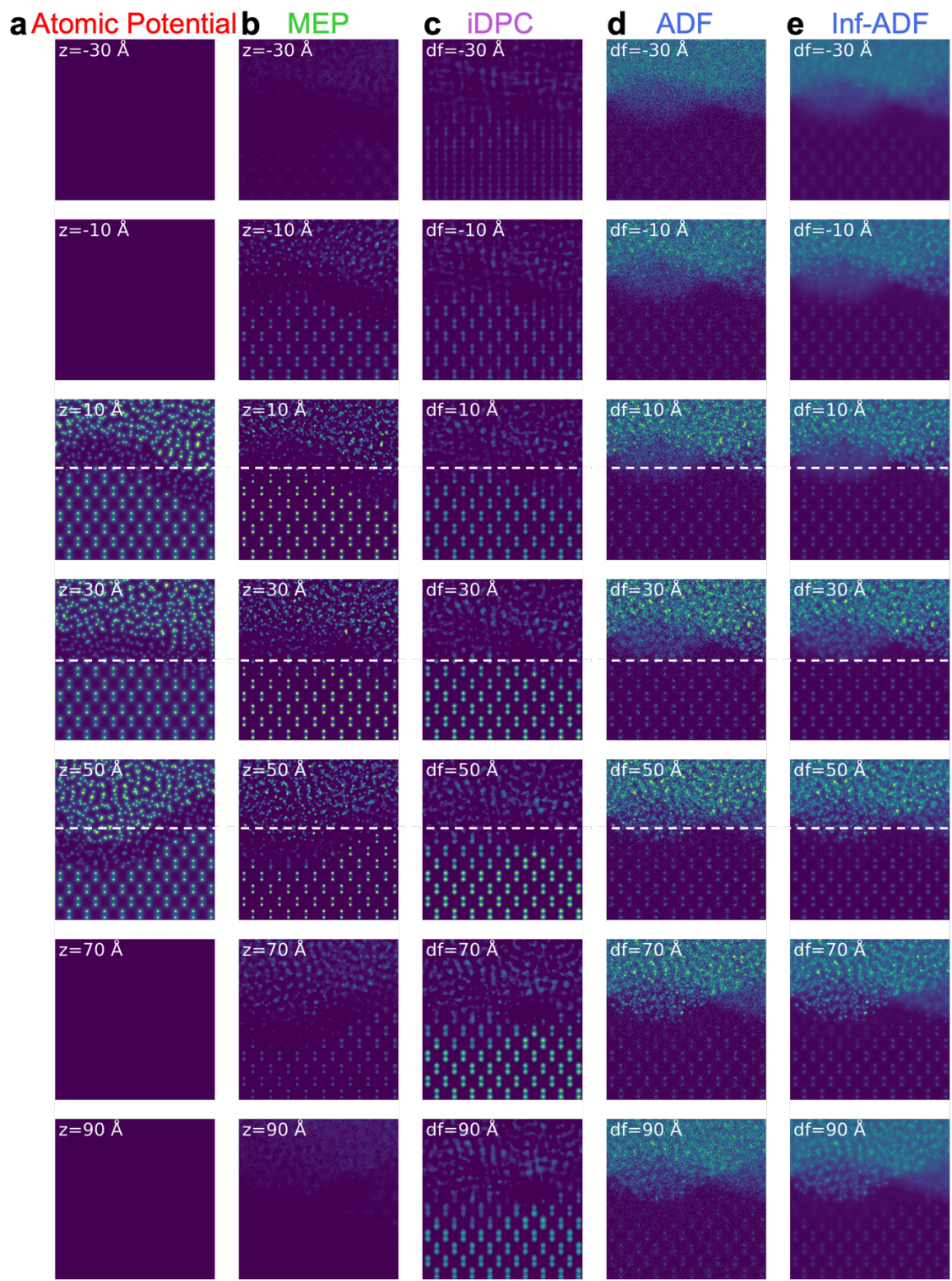

**Extended Data Figure 1. Selected depth slices showing depth-sectioning capabilities of MEP, tf-iDPC and tf-ADF imaging techniques for a c-Si/a-SiO$_2$/a-HfO$_2$ model interface. a** Ground

truth atomic potential depth slices, each 10 Å thick, with vacuum added before and after to match the total thickness of the simulations. **b** Corresponding depth slices of the MEP-reconstructed atomic potential, each 10 Å thick, accurately recovering the structure and roughness of the interface. **c** Corresponding tf-iDPC images from the stack, generated with a 10 Å defocus step, showing significant elongation of features in depth and loss of interface clarity. **d** tf-ADF images from the stack, also generated with a 10 Å defocus step, suffering from noise and poor depth resolution. **e** Corresponding infinite-dose (noise-free) tf-ADF images, highlighting the intrinsic elongation of features in depth even under ideal conditions. All images share a normalized intensity range from 0 to 1, scaled by the maximum value of their respective 3D stack. Spherical and chromatic aberrations are included in all datasets. Except for **e**, all having a total dose of $2.5 \times 10^5$ $e^-/Å^2$ per stack. The field of view is consistently 35 Å across all images. White dotted lines for z=10, 30, 50 Å are aligned with the topmost Si atom in the ground truth at the respective depths, highlighting the sinusoidal interface shape. Only the MEP reconstruction recovers the correct shape, depth position, and roughness of the interface.

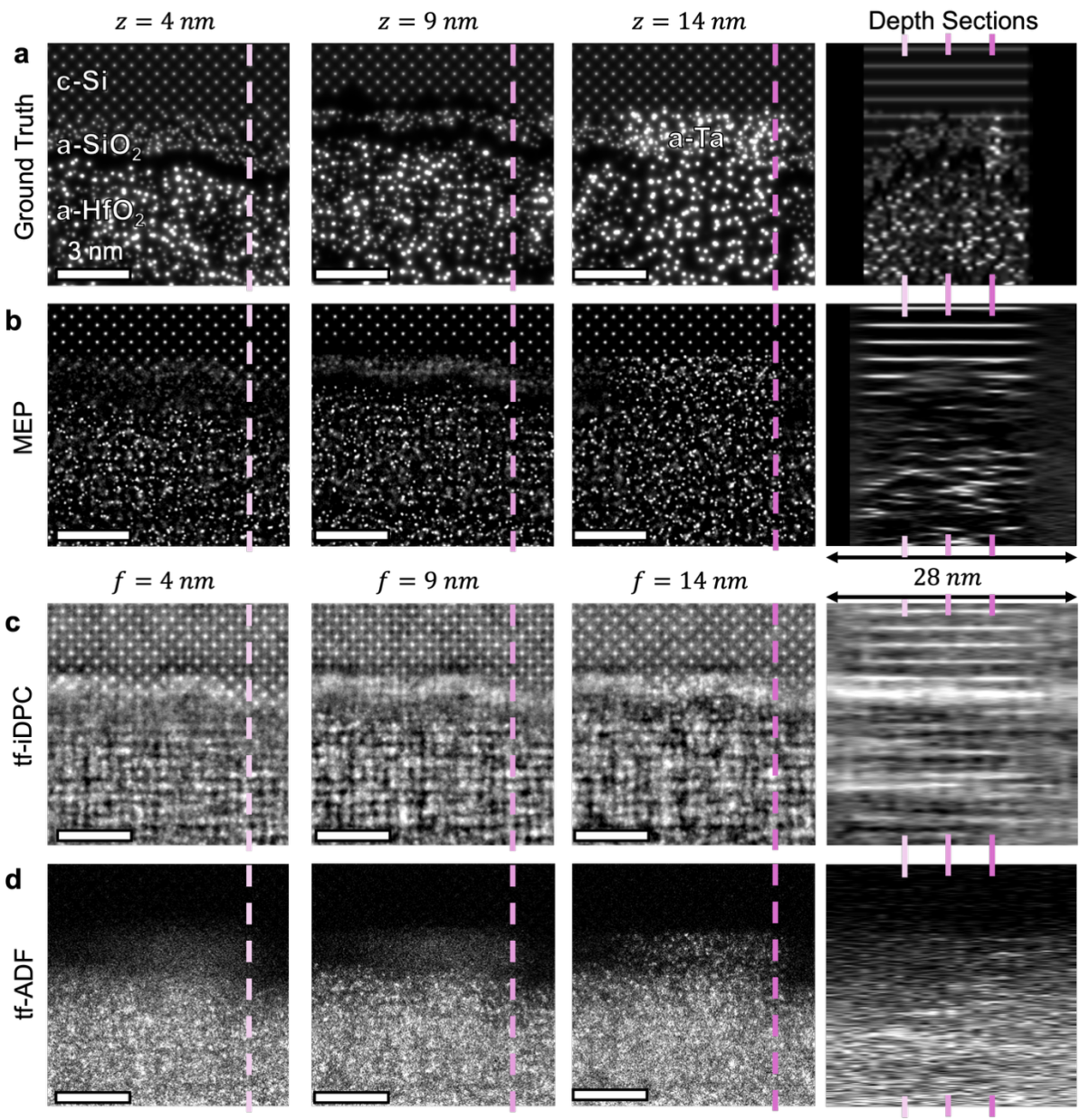

**Extended Data Figure 2. Benchmarking the 3D imaging capabilities of MEP, tf-iDPC, and tf-ADF through simulations of a pMOS device structure. a** Atomic potential of a section of a pMOS device structure with 1 nm depth blurring, shown at three depths. The model includes an intrusion defect filled with amorphous tantalum (Ta) (brighter region) at z = 14 nm. Dotted lines indicate the location along which the depth section is taken, shown in the rightmost panel. **b** MEP reconstruction atomic potential of the pMOS structure. The three panels correspond to the same depths as in **a**, with the intrusion defect resolved at z = 14 nm. Dotted lines indicate the depth section shown on the right. **c** Simulated tf-iDPC images at corresponding depths. The intrusion defect is difficult to resolve. **d** Simulated tf-ADF images at corresponding depths. While the intrusion defect is visible, the lighter atoms are not. The total dose for MEP and tf-iDPC simulated

stacks is 0.4 × 10$^5$ e$^-$/Å$^2$ and 1 × 10$^5$ e$^-$/Å$^2$ for tf-ADF stack (to have enough signal to see any features). All scale bars are 3 nm.

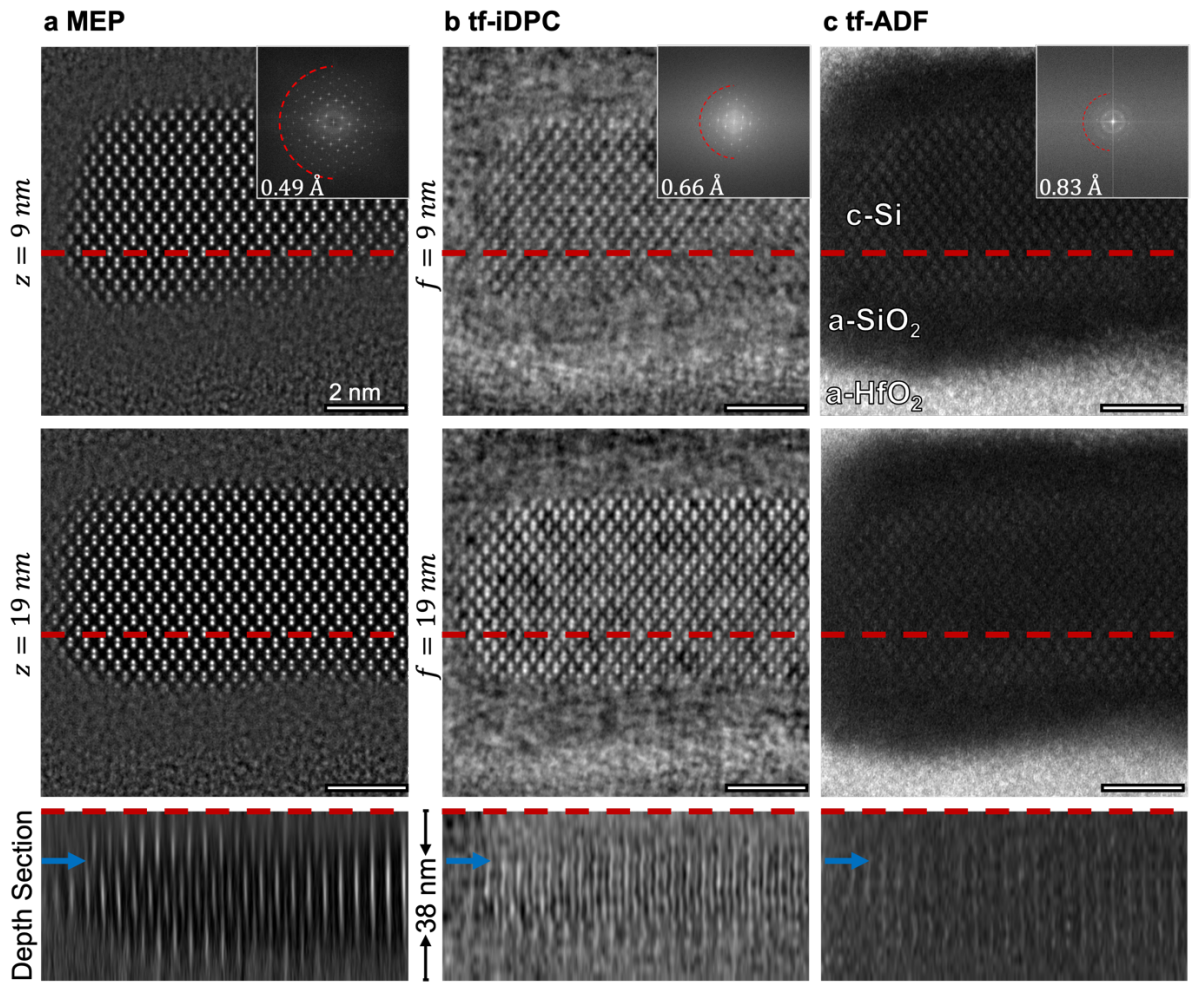

**Extended Data Figure 3. Selected slices comparing experimental measurements of the different methods and 2D information transfer, with a depth section highlighting a stacking fault. a** MEP, **b** tf-iDPC, and **b** tf-ADF images, each shown at two depths (9 nm and 19 nm), with a horizontal depth section (along the red dotted lines) below, illustrating stacking faults. MEP achieves higher in-plane resolution (0.49 Å) compared to tf-iDPC (0.66 Å) and tf-ADF (0.83 Å). Defects in depth reduce resolution and introduce artifacts in the through-focal imaging methods, complicating post-acquisition alignment, diminishing depth resolution, and reducing feature detection reliability. Stacking faults (blue arrows), clearly visible in MEP, are obscured in tf-iDPC and tf-ADF due to these artifacts. Total dose: 0.5 × 10$^5$ e$^-$/Å$^2$ for MEP, and 1 × 10$^5$ e$^-$/Å$^2$ for both tf-iDPC and tf-ADF.

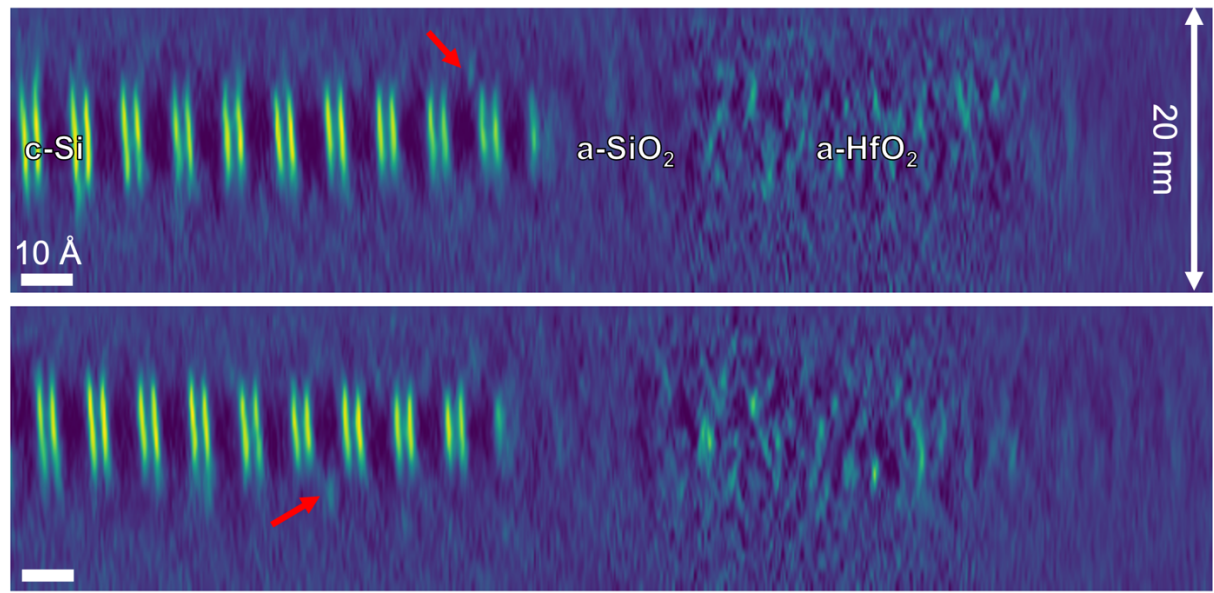

**Extended Data Figure 4. Experimental MEP depth sections from the planar c-Si/a-SiO$_2$/a-HfO$_2$ interface.** Depth sections reveal internal features of the amorphous HfO$_2$ layer, including nanoscale contrast variations consistent with short-range order or local density fluctuations. In contrast, the a-SiO$_2$ portion appears more homogeneous, consistent with its lower atomic number and more uniform amorphous structure. Heavy atoms located at the top and bottom surfaces of the lamella (indicated by red arrows) are identified as re-deposited from the TEM sample preparation process. The characteristic triangular geometry of the FIB-prepared TEM lamella is visible in the depth cross sections, confirming the wedge shape of the sample. These observations demonstrate the sensitivity of MEP to both intrinsic and preparation-induced structural features, highlighting its utility for detailed 3D characterization of amorphous-crystalline interfaces.

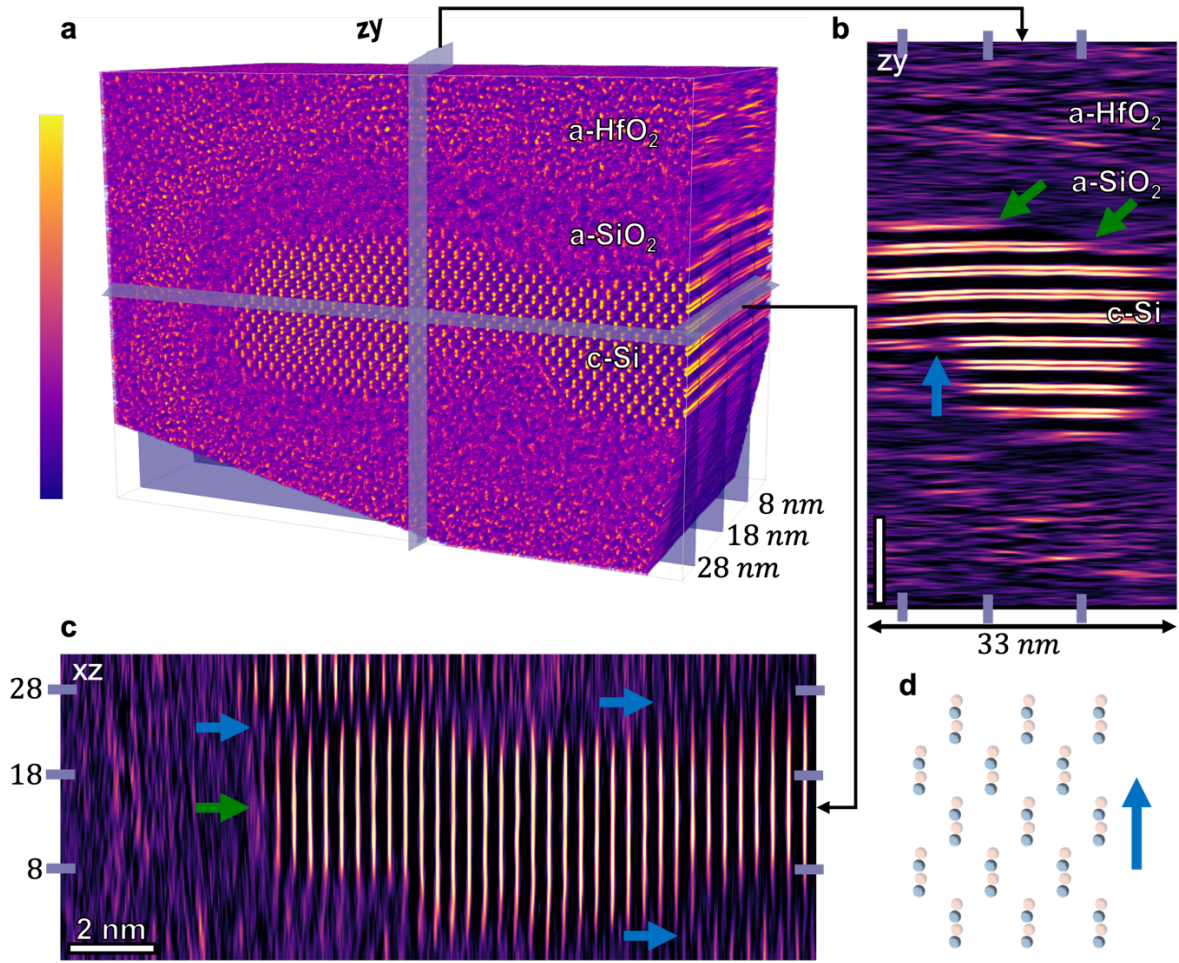

**Extended Data Figure 5. MEP reconstructed cuboid from measured data on the GAA device with depth sections revealing stacking faults and step edges. a** A 3D cuboid representation of the reconstructed device, illustrating all sliced planes (highlighted in green in panels **b** and **c**) with labeled components visible in three dimensions. The cuboid dimensions are 18 nm × 13 nm × 33 nm (not to scale in depth). **b** Depth section along the zy plane shown in **a**, highlighting a region with missing crystalline Si accompanied by a stacking fault (blue arrow). Step edges are visible at the c-Si/a-SiO$_2$ interface (green arrows), and bright streaks in the amorphous region indicate the presence of Hafnium atoms. **c** Depth section along the xz plane highlighted in **b**, showing additional stacking faults (blue arrows) and a step edge at the silicon interface (green arrow). **d** Schematic depiction of a stacking fault, illustrating a half-unit-cell mis-registry between Si layers in depth. Scale bars: 5 nm.

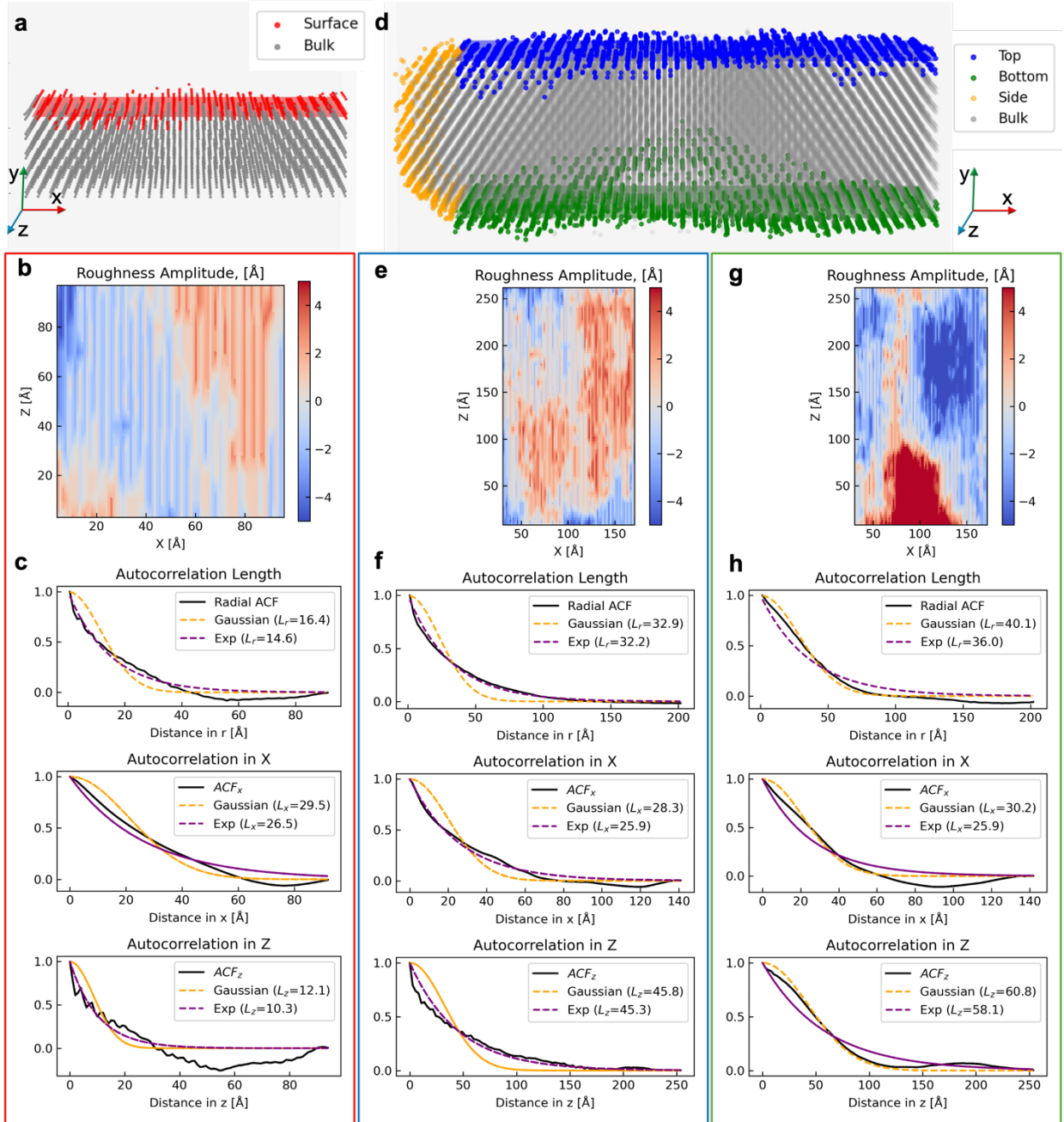

**Extended Data Figure 6: Atomic-scale roughness and its correlation length scales of planar and GAA c-Si/a-SiO$_2$ interfaces. a–c** Planar reference interface. **a** 3D rendering of tracked atoms: bulk atoms are shown in gray, while surface atoms and the fitted reference plane are shown in red. **b** Roughness amplitude as the residuals from the reference plane, representing local deviations form an atomically sharp interface. **c** Autocorrelation analysis of roughness: the radial autocorrelation function (ACF) (top) and line cuts along x and z (bottom). The planar interface exhibits an RMS roughness of 1.7 Å and a correlation length of ~15 Å. The ACF decays exponentially, consistent with abrupt, step-like morphology. **d–h** GAA 2 device. **d** 3D rendering of tracked atoms: bulk in gray; surface atoms and fitted facet planes colored top (blue) and bottom

(green). **e–f** Residuals and ACFs for the top interface, with RMS roughness of 2.1 Å and a correlation length of ~30 Å, again well-described by an exponential model. **g–h** Residuals and ACFs for the bottom interface, showing RMS roughness of 3.8 Å and a decay shape that deviates from both exponential and Gaussian models, suggesting more complex morphology (such as the "mouse bite"), likely driven by growth-related defects and interdiffusion (i.e. the "mouse-bite"). Across interfaces, exponential ACF models best describe smoother surfaces, while rougher ones show ambiguous behavior. This metrology reveals critical process-dependent variations relevant for classical scaling limits. These direct measurements can clarify longstanding ambiguity in interface roughness models derived from 2D projections and enable direct measurement of buried 3D morphology.

## Methods

### Device Fabrication

The GAA devices were fabricated following a customized version of the full electrical device flow developed by IMEC, with simplifications to enable direct access to critical structural features for characterization. Specifically, the process follows the basic CFET flow[60], modified to produce 3× nanosheet GAA structures using a different starting epitaxial stack. The middle dielectric isolation and Epi source/drain modules were skipped to allow early access to the nanosheet release and gate stack deposition steps. The devices were formed by first growing a superlattice of alternating Si/SiGe single-crystal epitaxial layers. The SiGe layers serve as sacrificial materials that are selectively etched to release the Si nanosheets, which ultimately act as the transistor channels. The resulting gaps are then used to deposit the gate stack (typically a-$SiO_2$/a-$HfO_2$/metal gate). An interfacial a-$SiO_2$ layer is first formed by oxidizing the Si nanosheets, followed by high-k and metal gate deposition via atomic layer deposition (ALD), which offers the precise thickness control and conformality needed to uniformly cover the nanosheets on all sides. This stacked geometry enables each ALD step to coat multiple Si channels simultaneously[3,60–62].

### TEM Sample Preparation

The c-Si/a-$SiO_2$/a-$HfO_2$ cross-sectional lamella was prepared using a standard cross-section Focused Ion Beam (FIB) lift-out procedure on Thermo Fisher Scientific Helios G4 UX FIB. The lamella is 10-20 nm thick in the region of interest. The GAA sample was prepared by Eurofins Nanolab Technologies with thickness of ~30-40 nm in the region of interest.

### Data Acquisition

All the experimental datasets were acquired using a Cs-corrected Thermo Fisher Scientific Spectra 300 X-CFEG at 300 keV with a probe convergence semi-angle of 30 mrad. The MEP datasets were acquired at 10-15 nm nominal overfocus values, using an EMPAD-G2 detector, with scan step sizes in real space of 0.45 – 0.66 Å, scan dwell time of 100 μs, and diffraction space sampling of 0.65 – 0.85 mrad/pixel. The scan fields of view were ~10-20 nm with 256x256 pixel scans. A total dose of $0.5 - 2.5 \times 10^5$ e$^-$/Å$^2$ was applied for the 4D-STEM datasets. The DPC images were acquired on the Thermo Fisher Scientific Panther detector (8 sections) and the ADF images were acquired using the Thermo Fisher Scientific ADF detector, with total through focal series dose of $1 \times 10^5$ e$^-$/Å$^2$ using an automated acquisition method outlined here[63]. Nominal defocus values were corrected by a factor of two based on calibration.

### MEP Reconstruction

Multislice electron ptychography (MEP) was employed to reconstruct the 3D atomic potential of the sample with sub-Ångström lateral and 3-4 nm depth resolution. MEP solves the inverse problem of structure retrieval using 4D-STEM datasets, where a coherent electron beam is raster-scanned across the sample, recording diffraction patterns at each scan position. These diffraction patterns encode both phase and amplitude information of the transmitted electron waves due to overlap in the diffraction spots, and encode depth information due to the parallax effect, allowing for depth-resolved structural reconstruction.

To model the sample, it is conceptually divided into thin slices perpendicular to the beam propagation direction, and phase retrieval is performed using the maximum-likelihood multislice ptychography algorithm implemented in the *fold_slice* package. This algorithm iteratively refines both the probe and sample structure, using a gradient-based approach and multiple probe modes to account for the partial coherence of the electron beam. The optimal reconstruction parameters – convergence semi-angle, defocus, and sample thickness – were determined using a Bayesian optimization model, minimizing data error as the objective function.

The reconstruction used 4 probe modes and >1000 iterations, with a slice thickness of 1 nm and depth regularization of 0.4–0.7 (reconstruction summaries in Supplementary Figure 1). Since each slice in the reconstruction represents a phase object, the amplitude of each slice should close to one, with deviations primarily due to scattering beyond the outer collection angle of the detector (thus hafnium oxide looks darker), as is shown in Supplementary Figure 1. High-pass filtering was applied to the final phase reconstructions to eliminate low-frequency artifacts. Supplementary Figure 1 provides a visual summary of the object phase and amplitude, with intensity histograms, as well as the different probe modes at the entrance surface for all the experimental datasets considered in this work. The ptychographic reconstructions were computed using an in-house cluster, requiring approximately one day for completion.

**Simulation of STEM and 4D-STEM Data**

To evaluate the imaging performance of multislice electron ptychography (MEP) and compare it with other imaging modes, we conducted multislice simulations of scanning transmission electron microscopy (STEM) images and 4D-STEM datasets. To validate the accuracy of MEP, we benchmarked its performance in reconstructing a technologically relevant c-Si/a-SiO$_2$/a-HfO$_2$ interface structure and a pMOS structure[12,54]. The simulations were performed using the *ab*TEM multislice simulation package[53]. The simulation approach provides a controlled environment for assessing resolution, sensitivity, and depth reconstruction accuracy while incorporating experimental conditions such as realistic aberrations and electron dose (through detector noise).

The simulations were conducted under conditions mimicking experimental setups, using a 300 keV acceleration voltage and an aberrated electron probe with a semi-convergence angle of 30 mrad. The simulated probe was subjected to spherical (Cs = 1 μm) and chromatic (Cc = 2.46 mm) aberrations with an energy spread of 0.4 eV. The chromatic broadening can be estimated as $\frac{dE}{E_0}C_c \approx 3.28$ nm, which is in the order of the expected depth resolution and does not affect the imaging modes drastically. On the other hand, spherical aberration shifts the effective entrance surface location by approximately $\sqrt{C_s\lambda} \approx 1.5$ nm, thus shifting the depth features primarily for the tf-ADF and tf-iDPC images. The total electron dose was kept constant at $2.5 \times 10^5$ e$^-$/Å$^2$ across all imaging methods to ensure a fair comparison and Poisson noise was applied to the diffraction patterns to replicate experimental conditions.

Simulated (4D-)STEM datasets were generated by real space scan step size of 0.14 Å for tf-iDPC and tf-ADF images and 0.44 Å for c-Si/a-SiO$_2$/a-HfO$_2$ MEP and 0.75 Å for GAA MEP. The extent of the pixelated detector was like experimental conditions with maximum collection angle of round 60 mrad and 128x128 diffraction patterns. Using these datasets, various imaging modes were calculated, including MEP, through focal integrated differential phase contrast (iDPC), and through focal annular dark field (tf-ADF) imaging, results of which are presented in Figure 3 and Extended Data Figures 1-2. Chromatic aberration was added by averaging simulations at different defocus values according to a gaussian focal spread. Ptychographic reconstructions of the 4D-STEM datasets were performed using the *fold_slice* package; iDPC images were generated from 4D-STEM datasets by computing DPCx and DPCy images, followed by Fourier integration. ADF images were simulated by integrating intensities in annular detectors. For iDPC, only the bright disk was used (30 mrad) while for ADF the inner and outer angles for integration were 60 – 150 mrad. Unlike ptychographic reconstruction, iDPC and ADF imaging require multiple through-focal datasets to approximate 3D structural information, increasing the total dose and complexity.

The reconstructed images were compared to the ground truth atomic potential, demonstrating that MEP successfully recovered key structural features, including interface roughness, amorphous intrusions, and Si atomic column extend (see Figure 3 and Extended Data Figures 1 and 2). These features are critical for evaluating the performance and reliability of GAA transistors. Compared to iDPC and ADF, MEP provided superior 3D reconstruction accuracy and dose efficiency, mitigating artifacts associated with multiple scattering and channeling.

**Atom Tracking and Strain Relaxation Analysis**

Atom positions were identified using the Atomap package[55] (see Supplementary Figure 2 and Supplementary Movies 1,2 and 4). Custom-written Python scripts were used to segment tracked atoms into layers and analyze interfacial structures (see Supplementary Figure 2 and Supplementary Movies 1,2 and 4). Si-Si projected nearest neighbor distances were calculated for each layer for all the depths for all 3 datasets discussed in this work (see Figure 6c). Distance mappings were generated to visualize variations in bonding configurations across the layers in each sample (see Figure 6b).

**3D Visualizations**

Tomviz was used for volumetric visualization of atomic cuboids shown in Figure 1d, Figure 5b and Extended Data Figure 4a[64]. The c-Si/a-SiO$_2$ interface shape was constructed using the outer-most tracked Si atoms per layer and a 3D surface was constructed using Open3D python package, results of which are shown in Figure 6d-e, Extended Data Figure 5a,d and Supplementary Movies 3 and 5[65].

**Interface Roughness Quantification and Autocorrelation Analysis**

To quantify interface morphology, we compute standard roughness metrics—the root-mean-square (RMS) height and in-plane correlation length—directly from the tracked 3D atomic positions. For each dataset, the outermost interfacial Si atoms were segmented into facet-specific groups: planar, top-GAA, and bottom-GAA interfaces. A reference plane was first fitted to the local atomic surface for each facet using a least-squares approach. Height residuals were then computed as the perpendicular distances of each atom from this fitted plane.

The RMS roughness was defined as the average amplitude of deviation of these residual heights across each facet. To assess the lateral spatial coherence of the roughness, we computed the 2D autocorrelation function (ACF) of the roughness map. The in-plane correlation length, $L_c$, was extracted by fitting the ACF envelope with both exponential and Gaussian models:

Exponential decay: $C(r) \propto \exp(-r/L_c)$

Gaussian decay: $C(r) \propto \exp(-r^2/L_c^2)$

Both fitting models are presented (Extended Data Fig. 5h–i), as there is longstanding discussion in the literature about which model more accurately represents interfacial roughness. Since our MEP reconstructions yield fully 3D atom-resolved data, we can compute the autocorrelation either radially or along specific in-plane directions (x and z), shown in Extended Data Figure 5. This enables an analysis of anisotropy in the roughness correlation. Axis-resolved correlation lengths are reported alongside radially averaged values to capture the directional dependence of morphological fluctuations.

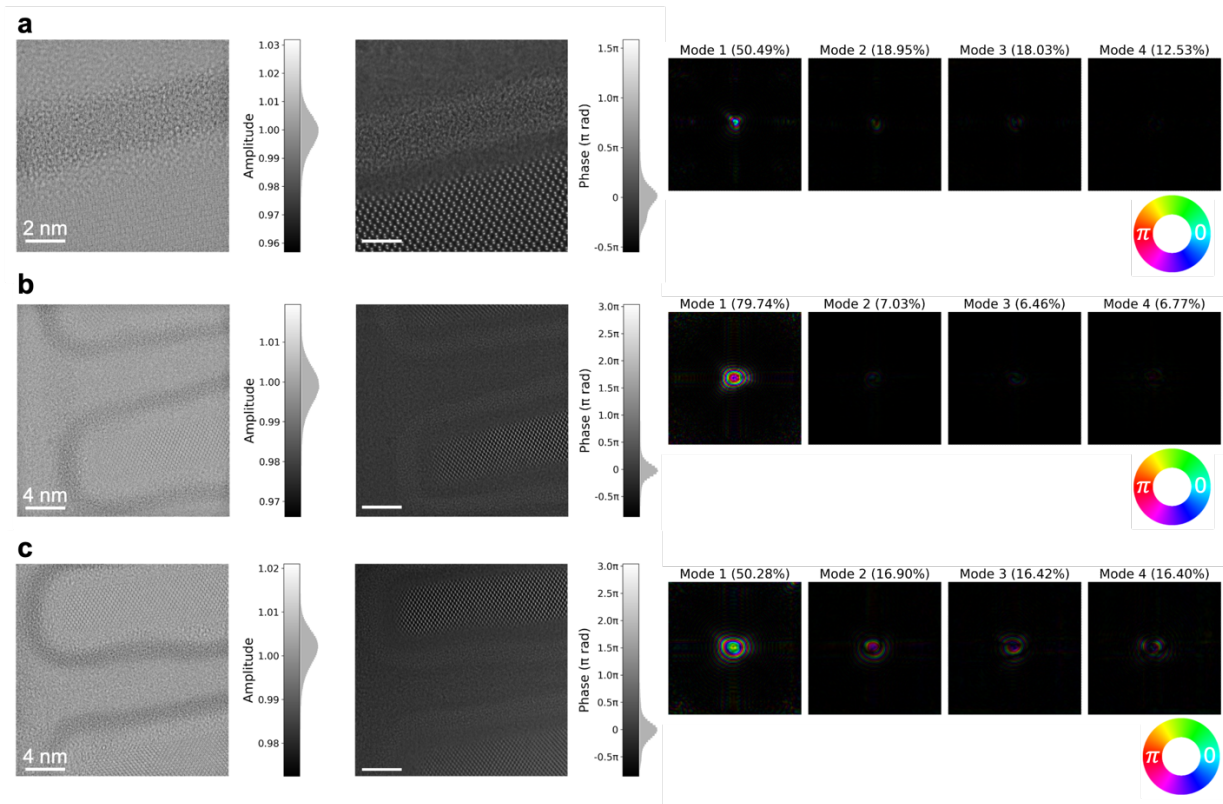

**Supplementary Figure 1: MEP reconstruction summaries: a planar interface, b GAA structure 1, and c GAA structure 2.** For each row, from left to right: stack-averaged amplitude, projected phase of the reconstructed object (both with corresponding color bars), and histograms showing the distributions of these values. Also shown are the reconstructed mixed-state probe amplitude (brightness) and phase (color) distributions at the entrance surface for each probe mode, along with their respective intensity percentages. The plotted object amplitude represents the average across all slices, while the projected object phase corresponds to the cumulative phase contributions from all slices. The amplitude for each slice remains close to one, as each slice behaves like a near-pure phase object. Deviations from unity are primarily due to scattering beyond the outer detector collection angle, which is more pronounced in regions containing heavier scatterers, such as hafnium (Hf). Residual aberrations, including the trifold astigmatism visible in the reconstructed probes, underscore the robustness of the MEP algorithm in recovering and separating both object and probe characteristics, even under complex experimental conditions.

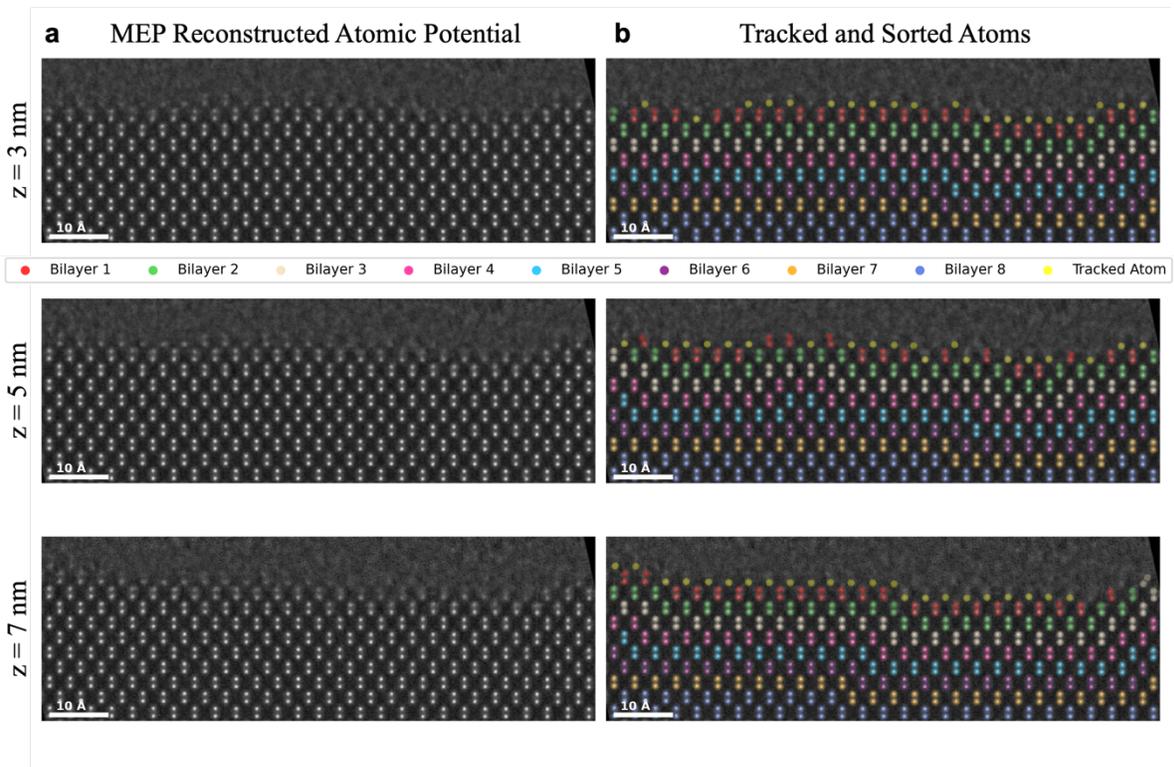

**Supplementary Figure 2: Selected slices of atom tracking on the measured planar c-Si/a-SiO$_2$ interface. a** Selected depth slices of MEP reconstructed atomic potential as a function of depth. **b** Same as **a** with superimposed tracked atoms separated into Si-Si bilayers from the interface. Tracked atoms without a nearest neighbor are colored yellow.

**Supplementary Movie 1**: Reconstructed 3D potential of the planar c-Si/a-SiO$_2$ interface. **a.** Multislice Electron Ptychography (MEP) reconstructed electrostatic potential visualized as a function of depth. **b.** Same as **a**, with tracked atomic positions overlaid and grouped into bilayers from the interface.

**Supplementary Movie 2:** Reconstructed 3D potential of GAA Device 1. **a.** MEP reconstructed electrostatic potential as a function of depth. **b.** Same as **a**, with tracked atoms overlaid and organized into bilayers from the gate-channel interface.

**Supplementary Movie 3:** Reconstructed 3D potential of GAA Device 2. **a.** MEP reconstructed electrostatic potential as a function of depth. **b.** Same as **a**, with tracked atoms overlaid and separated into bilayers from the interface.

**Supplementary Movie 4**: 3D visualization of the silicon channel geometry in GAA Device 1. The reconstructed shape of the crystalline silicon channel is shown in 3D, highlighting variations in cross-sectional geometry and showing the various missing crystalline regions.

**Supplementary Movie 5**: 3D visualization of the silicon channel geometry in GAA Device 2. Same as Movie 4, but for GAA Device 2, which has more regular shape in comparison.

**Supplementary Movie 6**: Summary of MEP reconstruction of GAA Device 1. Sequential cross-sectional views through the 3D MEP reconstruction, highlighting the crystalline silicon channel, the "mouse-bite", and its extracted surface geometry.